\documentclass[%
 aip,
 amsmath,amssymb,
 reprint,%
]{revtex4-1}

\usepackage{graphicx}
\usepackage{dcolumn}
\usepackage{bm}

\usepackage[utf8]{inputenc}
\usepackage[T1]{fontenc}
\usepackage{mathptmx}

\begin{document}

\preprint{AIP/123-QED}

\title[16-Element Superconducting Nanowire Single-Photon Detector for Gigahertz Counting at 1550-nm]{16-Element Superconducting Nanowire Single-Photon Detector for Gigahertz Counting at 1550-nm}

\author{Timothy M. Rambo}
 \email{tim@quantumopus.com.}
 \affiliation{Quantum Opus LLC. Novi, Mi}
\author{Amy. R. Conover}%
 
\affiliation{Quantum Opus LLC. Novi, Mi}%

\author{Aaron J. Miller}
\affiliation{Quantum Opus LLC. Novi, Mi}%

\date{\today}

\begin{abstract}
We present a linearly arrayed, 16-element, superconducting nanowire single-photon detector with 83.4$\%$ system detection efficiency at 1550 nm and a mean per-element dead-time of 9.6-ns \textemdash enabling counting at 1 giga-count per second with $>50\%$ System Detection Efficiency. This device was designed and fabricated in an existing scalable commercial process. 
\end{abstract}

\maketitle

%

\section{\label{sec:intro}Introduction}

Superconducting nanowire single-photon detectors (SNSPDs) have proliferated the optics community in recent years due to their unique combination of high detection efficiency ($\geq 98 \% $)\cite{NIST98pct}, extremely low dark counts ($<1/$s for UV\cite{JPLUV}, $<100/$s commercially avilable at 1550 nm), broad spectral response from UV to mid-infrared\cite{JPLUV,SNSPDMIRSPEC,Hadfield2p3umLIDAR}, precise timing accuracy ($<$4.3 ps FWHM intrinsic timing jitter at 1550 nm for NbN nanowires, and $<$10.6 ps FWHM intrinsic timing jitter at 1550 nm for MoSi)\cite{IntrinsicNbNJitter,IntrinsicMolyJitter}, and recent discovery of true photon number resolving capacity\cite{OpticaPNR,STANDPNR}. Diverse fields as quantum optics\cite{ThreePhoton,20ModeBosonSampling,KwiatQD,GauthierQKD}, spectroscopy\cite{AtallahSpectroscopy}, and free-space optical communications\cite{LLCDsummary,GRCsnspd} have leveraged SNSPDs with great success and driven the rise of commercially available SNSPD systems. Commercial SNSPDs are typically single-element devices which can achieve count rates $\sim$10 MHz and have small active areas designed for coupling to single-mode fibers. 
To keep up with the evolving demands of the research community, SNSPD technology must increase it’s achievable count-rates and active areas. While it is possible to change the electrical characteristics by adding series resistance to achieve higher count rates\cite{KermanET,QOSPW} and increase the footprint of SNSPDs\cite{NIST98pct}, meandering the wires over a larger area increases the kinetic inductance which slows recovery time. One approach to circumvent this counteraction is to produce multi-element SNSPDs, where the device is fabricated to cover a larger area but split into many active areas instrumented either indpendently\cite{Huang_2018} or jointly\cite{NISTkilopixel,ThermalArray}. Recent work has also shown the viability of detectors made of out micron-scale superconducting wires with thinner superconducting films\cite{microwire}. Here, we present a linearly-arrayed, 16-element, SNSPD which achieves 83.4$\%$ system detection efficiency (SDE) and is capable of counting with $\geq 50 \%$ at $\sim$1 GHz countrate. This device was developed in a commercial process for immediately scalable production.

\section{\label{sec:fab}Fabrication}

\begin{figure}[h]
    \centering
    \includegraphics{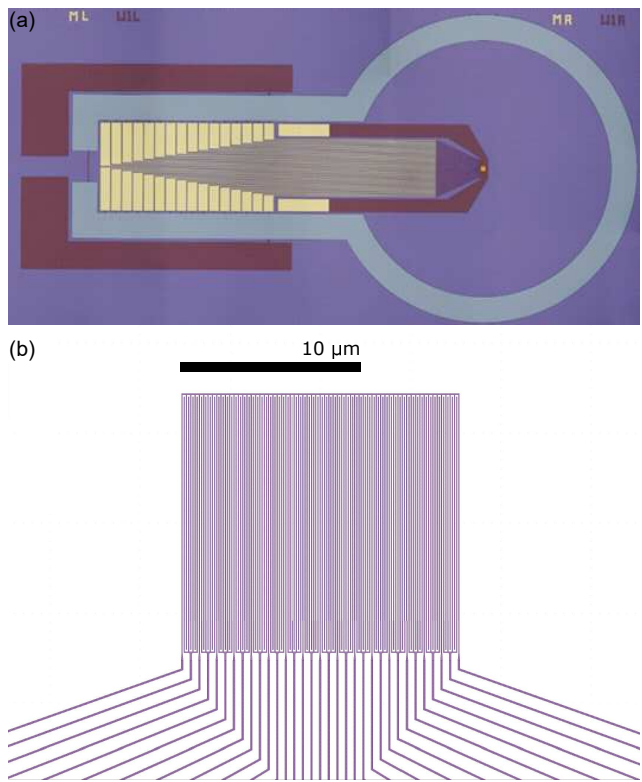}
    \caption{(a) An optical micrograph showing the structure of the multi-element SNSPD chip. The round region at the end is 2.5mm in diameter for compatibility with zirconia sleeves for self-alignment to fiber and the SNSPD is too small to see. (b) Zoomed in view of the multi-element detector. This is a negative image, white shows the presence of superconductor and purple the absence.}
    \label{fig:multielement}
\end{figure}

Our device was fabricated on a Silicon substrate topped with a $\sim$100-nm thick thermal oxide layer beneath a gold mirror. An additional oxide layer separated the device and the mirror, creating the back-short for a $\lambda/4$ optical cavity formed around the device. The SNSPD itself was fabricated by depositing a 7 nm thick proprietary metal and amorphous sillicide film, with a 5 K critical temperature on top of the back-short. We lithographically patterned this film into a meandered wire geometry with 60 nm wire-width and 50$\%$ fill factor to create 16 linearly arrayed, individually wired, 16 $\mu$m $\times$ 1 $\mu$m, sub-elements to create the SNSPD device. The top side of the $\lambda/4$ cavity was created using a two-layer dielectric coating to enhance absorption at 1550 nm. As a final step to the fabrication process, the wafer was deep etched to  create a keyhole like chip for the detector compatible with the self-alignment technique\cite{SelfAlignment}. An optical micrograph of the device chip and an up-close view of the linearly arrayed, independently wired, elements is shown in Figure \ref{fig:multielement}.

\section{\label{sec:char}Characterization}
The device was characterized at a temperature of 2.3K in an Opus One rack-mountable Gifford-McMahon cryostat with illumination at 1550 nm provided via self-alignment to OFS $\#$36484 four-mode graded index fiber (FMF). The detector was electrically instrumented using a patent pending, dc-coupling, anti-latch circuit\cite{QOpatent} \textemdash which requires a single-coax line per element as opposed to the two lines required for a typical cryogenic bias tee\textemdash and the commercially available Quantum Opus SNSPD Bias and Amplification Module, a self-contained 1U system. To characterize SDE, a tunable continuous wave laser was injected into the FMF from singlemode fiber (SMF) via a series of cascaded attenuators. Incident photon flux was set to $\sim3 \times 10^{6}/$s for the measurement of SDE as a function of device bias for each element. Net SDE across all elements was 83.4$\%$ with 71,000 dark counts per second. FMF laser sources are not readily available, but there has been an  exploration\cite{GRCFMF} of how the higher order modes in the fiber couple to an SNSPD with similar total footprint to this one, and how the increased fiber core size affects dark rates compared to SMF coupling. 

\begin{figure}
    \centering
    \includegraphics{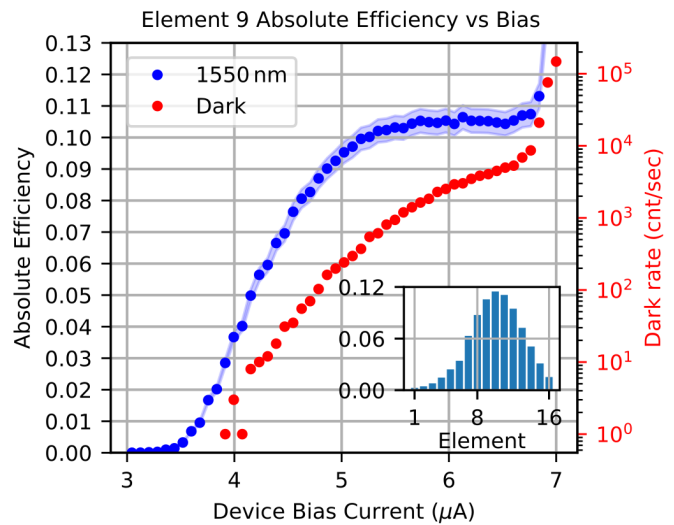}
    \caption{ Absolute efficiency and dark counts for element 9 of the SNSPD as a function of bias current. (inset) Efficiency at the bias current which yields 5,000 dark for each element.}
    \label{fig:efficiency}
\end{figure}

Black body radiation is the dominant source of noise in this experiment. The flux of black body photons at room temperature tends to become significant for wavelengths $>$1700 nm which are very poorly guided by SMF, typically creating 100 dark counts per second on commercially available SNSPDs optimized for 1550nm. However, the FMF in this experiment supports multiple modes at wavelengths even beyond 2 $\mu$m resulting in an elevated dark rate which would not be present in all applications or for all fiber types. The dark counts are further enhanced by the fact that the elements in this device have a relatively long plateau of saturated 1550-nm detection efficiency as a function of bias currents. A long plateau is suggestive of a better response for the blackbody photons than for devices which barely reach saturation. Figure \ref{fig:efficiency} shows a system detection efficiency and dark count rate versus bias plot for one of the elements. Charting the efficiency as a function of element number shows that the peak photon absorption occurs off center of the sub-element array. This is due to alignment uncertainty in the fabrication process and fiber tolerances which cause an offset between the center of the keyhole structure and the SNSPD.

\begin{figure}
    \centering
    \includegraphics{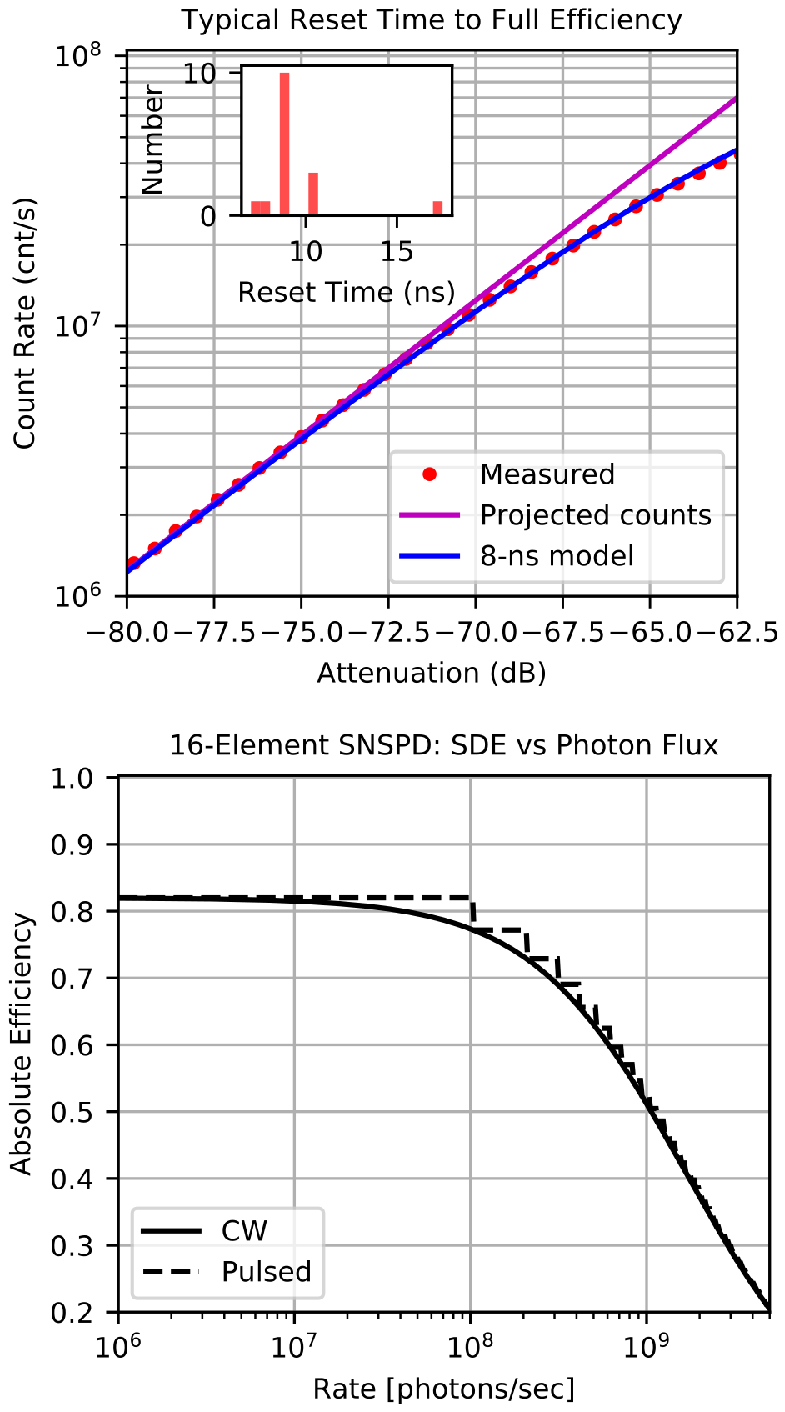}
    \caption{(top) Count-rate versus source attenuation measurement for SNSPD element 12. Modeled versus projected counts show 8-ns reset time. Inset shows histogram of reset times across all elements. Histogram of full-efficiency reset times of the sub-elements in the detector and calculated net detection efficiency versus incident photon flux from a CW source. (bottom) Model of SDE versus incident photon flux based on measured reset times of elements assuming unifrom illumination. SDE is $\geq 50\%$ for count rates up to 1 GHz.}
    \label{fig:countrate}
\end{figure}

To measure reset-time to full efficiency, we swept incident photon rates through the range of $\sim10^{3}$/s to $\sim10^{8}$/s with the SNSPDs biased to the current which gives 5,000 dark counts per second. The resulting counts versus attenuation data was then fit using a Poisson model to determine the reset-time. Figure \ref{fig:countrate} shows data and the Poisson modelling for a single element as well as a summary of data for all elements. Measurements across all sub-elements of the SNSPD yielded an average rest-time to full efficiency of 9.6 ns with 2.2 ns standard deviation, corresponding to an average maximum full-efficiency pulsed count rate of 107.6 MHz. Using these numbers, we calculate efficiency as a function of incident photon rate for the entire device, and this shows the device is capable of counting at $\geq$ 50$\%$ SDE with $\sim$1 GHz incident photon rate under uniform illumination.

Jitter was measured using a 1550-nm, mode-locked fiber laser (Nuphoton EDFL-PICO-1560-5-50-FCA) and a Swabian Time Tagger Ultra. Each element was biased to a current which corresponded to 5,000 dark counts per second for the jitter measurement. Jitter was measured as a correlation between the laser electrical reference and the electrical pulse output from each SNSPD element. Systemic sources of jitter were the intrinsic jitter of the Time Tagger Ultra, ,$J_{TT}=22$ ps FWHM, and the optical pulse width of the laser, $J_{pulse}= 8$ ps FWHM. Jitter contribution form the electrical reference was determined to be negligible compared to the other sources. The FWHM jitter contribution of the SNSPD, $J_{SNSPD}$, was extracted from the measured FWHM jitter, $J_{meas}$, using the following relation:
\begin{equation}
    J_{meas}^{2}=J_{SNSPD}^{2}+J_{TT}^2+J_{pulse}^2.
    \label{eq:jitter}
\end{equation}
A typical timing response histogram can be seen in Figure\ref{eq:jitter}. Mean jitter across all elements was 86.1 ps with a standard deviation of 10.2 ps.

One element had significantly worse jitter than the others. It also had a reduced switching current, no detection efficiency plateau, and reduced count-rate compared to the other elements. This suggests a constriction or defect in this one element. In a standard, single-element, SNSPD design this defect would have limited the performance of an entire device, but here affects the performance of only 1/16th of the device.

\begin{figure}
    \centering
    \includegraphics{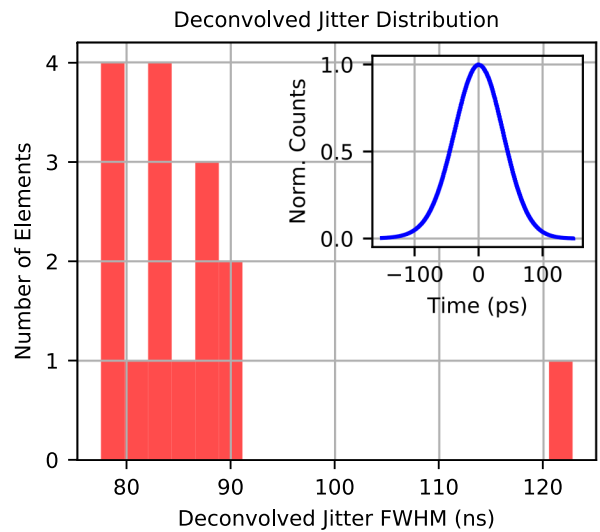}
    \caption{(Distribution of deconvolved jitters across all elements. Inset shows a timing distribution from a single element.}
    \label{fig:jitter}
\end{figure}

\begin{figure*}[ht]
    \centering
    \includegraphics{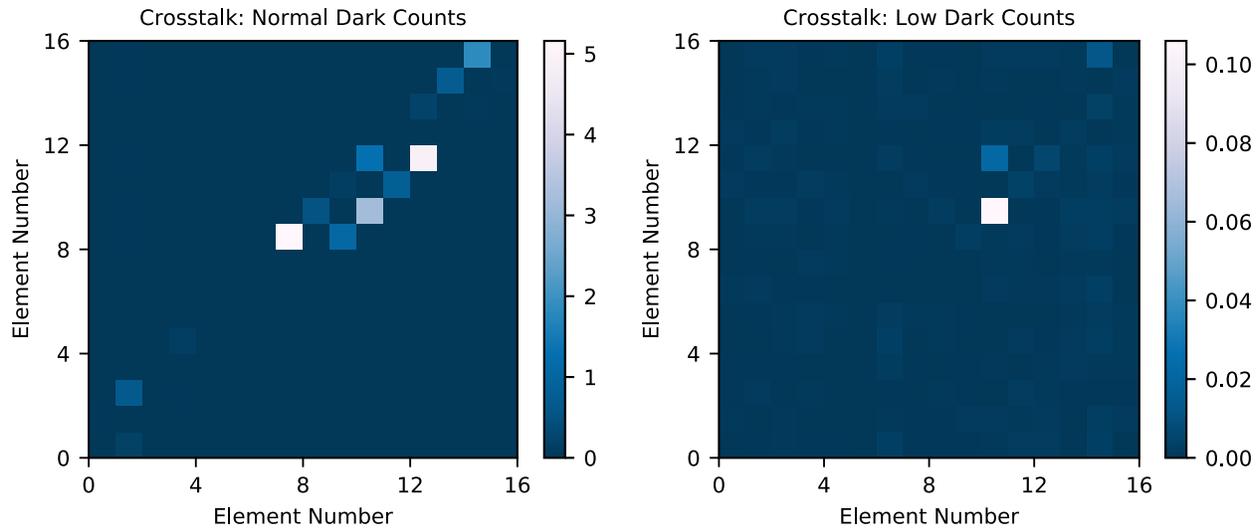}
    \caption{(left) Crosstalk percent for each element pair when one channel is biased to 5,000 dark counts per second and the other to 100,000 dark counts per second. (right) Crosstalk percent for each element pair when one channel is biased to 500 dark counts per second and the other to 100,000 dark counts per second}
    \label{fig:xtalk}
\end{figure*}

Element-to-element crosstalk was measured pair-wise in all possible combinations. This was done by biasing a given pair of elements $E_i$ and $E_j$ to the 5,000 dark count bias, then biasing $E_i$ into a high dark count regime $\sim 10^6$ dark counts/s and measuring the increase in counts on $E_j$ as a percent of the total count rate on $E_i$. The maximum observed increase in counts for any $E_j$ was $5\%$ of the counts on $E_i$, with most $i,j$ combinations yielding $<< 1\%$ increase. When repeated with starting bias currents which corresponded to 500 dark counts per second on each channel, the maximum extra counts observed on any $E_i$ was $0.1\%$ with the majority of $i,j$ combinations giving $<<0.02\%$, nearly an order of magnitude improvement. Because the cross-talk of a handful of element pairs dominates the others, bias adjustment would only be necessary on a subset of elements to suppress the effect. This would help to mitigate the efficiency penalty of lowering bias currents. A full plot of the results from both bias regimes can be seen in Figure \ref{fig:xtalk}. Using channel $E_i$ as the start channel on a time-tagger, we observed that the crosstalk counts on stop channel $E_j$ were delayed by $\sim$1 ns. This suggests that it may also be possible to remove or reduce the cross-talk using temporal gating.

\section{\label{sec:conc}Conclusion}
We presented an FMF coupled 16-element SNSPD created from an existing, immediately scalable, commercial process. This multi-element device showed $83.4\%$ net system detection efficiency with 71,000 dark counts per second, average per-pixel timing jitter of 86.1 ps FWHM, and average per-pixel reset time of 9.675 ns\textemdash meaning the device can count with $>50\%$ system detection efficiency at 1GHz mean photon rate. Dark counts are elevated above standard commercial SNSPDs due to the FMF-coupling, these could be drastically reduced by coupling the device to SMF (which achieves dark count rates of <100 /s on standard commercial devices) or by using cold filtering techniques \cite{MMsnspdFiltering}. We expect that increasing availability of this type of high-count rate, high efficiency, SNSPD will drive new applications in many fields.
\newpage

\begin{acknowledgments}
We wish to acknowledge the support of NASA grant 80NSSC19C0141 for helping to make this research possible.
\end{acknowledgments}

\nocite{*}
\bibliography{16elementFMF}

\providecommand{\noopsort}[1]{}\providecommand{\singleletter}[1]{#1}%
\begin{thebibliography}{26}%
\makeatletter
\providecommand \@ifxundefined [1]{%
 \@ifx{#1\undefined}
}%
\providecommand \@ifnum [1]{%
 \ifnum #1\expandafter \@firstoftwo
 \else \expandafter \@secondoftwo
 \fi
}%
\providecommand \@ifx [1]{%
 \ifx #1\expandafter \@firstoftwo
 \else \expandafter \@secondoftwo
 \fi
}%
\providecommand \natexlab [1]{#1}%
\providecommand \enquote  [1]{``#1''}%
\providecommand \bibnamefont  [1]{#1}%
\providecommand \bibfnamefont [1]{#1}%
\providecommand \citenamefont [1]{#1}%
\providecommand \href@noop [0]{\@secondoftwo}%
\providecommand \href [0]{\begingroup \@sanitize@url \@href}%
\providecommand \@href[1]{\@@startlink{#1}\@@href}%
\providecommand \@@href[1]{\endgroup#1\@@endlink}%
\providecommand \@sanitize@url [0]{\catcode `\\12\catcode `\$12\catcode
  `\&12\catcode `\#12\catcode `\^12\catcode `\_12\catcode `\%12\relax}%
\providecommand \@@startlink[1]{}%
\providecommand \@@endlink[0]{}%
\providecommand \url  [0]{\begingroup\@sanitize@url \@url }%
\providecommand \@url [1]{\endgroup\@href {#1}{\urlprefix }}%
\providecommand \urlprefix  [0]{URL }%
\providecommand \Eprint [0]{\href }%
\providecommand \doibase [0]{http://dx.doi.org/}%
\providecommand \selectlanguage [0]{\@gobble}%
\providecommand \bibinfo  [0]{\@secondoftwo}%
\providecommand \bibfield  [0]{\@secondoftwo}%
\providecommand \translation [1]{[#1]}%
\providecommand \BibitemOpen [0]{}%
\providecommand \bibitemStop [0]{}%
\providecommand \bibitemNoStop [0]{.\EOS\space}%
\providecommand \EOS [0]{\spacefactor3000\relax}%
\providecommand \BibitemShut  [1]{\csname bibitem#1\endcsname}%
\let\auto@bib@innerbib\@empty
\bibitem [{\citenamefont {Reddy}\ \emph
  {et~al.}(2019{\natexlab{a}})\citenamefont {Reddy}, \citenamefont {Lita},
  \citenamefont {Nam}, \citenamefont {Mirin},\ and\ \citenamefont
  {Verma}}]{NIST98pct}%
  \BibitemOpen
  \bibfield  {author} {\bibinfo {author} {\bibfnamefont {D.~V.}\ \bibnamefont
  {Reddy}}, \bibinfo {author} {\bibfnamefont {A.~E.}\ \bibnamefont {Lita}},
  \bibinfo {author} {\bibfnamefont {S.~W.}\ \bibnamefont {Nam}}, \bibinfo
  {author} {\bibfnamefont {R.~P.}\ \bibnamefont {Mirin}}, \ and\ \bibinfo
  {author} {\bibfnamefont {V.~B.}\ \bibnamefont {Verma}},\ }\bibfield  {title}
  {\enquote {\bibinfo {title} {Achieving 98\% system efficiency at 1550 nm in
  superconducting nanowire single photon detectors},}\ }in\ \href {\doibase
  10.1364/CQO.2019.W2B.2} {\emph {\bibinfo {booktitle} {Rochester Conference on
  Coherence and Quantum Optics (CQO-11)}}}\ (\bibinfo  {publisher} {Optical
  Society of America},\ \bibinfo {year} {2019})\ p.\ \bibinfo {pages}
  {W2B.2}\BibitemShut {NoStop}%
\bibitem [{\citenamefont {Wollman}\ \emph {et~al.}(2017)\citenamefont
  {Wollman}, \citenamefont {Verma}, \citenamefont {Beyer}, \citenamefont
  {Briggs}, \citenamefont {Korzh}, \citenamefont {Allmaras}, \citenamefont
  {Marsili}, \citenamefont {Lita}, \citenamefont {Mirin}, \citenamefont {Nam},\
  and\ \citenamefont {Shaw}}]{JPLUV}%
  \BibitemOpen
  \bibfield  {author} {\bibinfo {author} {\bibfnamefont {E.~E.}\ \bibnamefont
  {Wollman}}, \bibinfo {author} {\bibfnamefont {V.~B.}\ \bibnamefont {Verma}},
  \bibinfo {author} {\bibfnamefont {A.~D.}\ \bibnamefont {Beyer}}, \bibinfo
  {author} {\bibfnamefont {R.~M.}\ \bibnamefont {Briggs}}, \bibinfo {author}
  {\bibfnamefont {B.}~\bibnamefont {Korzh}}, \bibinfo {author} {\bibfnamefont
  {J.~P.}\ \bibnamefont {Allmaras}}, \bibinfo {author} {\bibfnamefont
  {F.}~\bibnamefont {Marsili}}, \bibinfo {author} {\bibfnamefont {A.~E.}\
  \bibnamefont {Lita}}, \bibinfo {author} {\bibfnamefont {R.~P.}\ \bibnamefont
  {Mirin}}, \bibinfo {author} {\bibfnamefont {S.~W.}\ \bibnamefont {Nam}}, \
  and\ \bibinfo {author} {\bibfnamefont {M.~D.}\ \bibnamefont {Shaw}},\
  }\bibfield  {title} {\enquote {\bibinfo {title} {Uv superconducting nanowire
  single-photon detectors with high efficiency, low noise, and 4 k operating
  temperature},}\ }\href {\doibase 10.1364/OE.25.026792} {\bibfield  {journal}
  {\bibinfo  {journal} {Opt. Express}\ }\textbf {\bibinfo {volume} {25}},\
  \bibinfo {pages} {26792--26801} (\bibinfo {year} {2017})}\BibitemShut
  {NoStop}%
\bibitem [{\citenamefont {Chen}\ \emph {et~al.}(2018)\citenamefont {Chen},
  \citenamefont {Schwarzer}, \citenamefont {Lau}, \citenamefont {Verma},
  \citenamefont {Stevens}, \citenamefont {Marsili}, \citenamefont {Mirin},
  \citenamefont {Nam},\ and\ \citenamefont {Wodtke}}]{SNSPDMIRSPEC}%
  \BibitemOpen
  \bibfield  {author} {\bibinfo {author} {\bibfnamefont {L.}~\bibnamefont
  {Chen}}, \bibinfo {author} {\bibfnamefont {D.}~\bibnamefont {Schwarzer}},
  \bibinfo {author} {\bibfnamefont {J.~A.}\ \bibnamefont {Lau}}, \bibinfo
  {author} {\bibfnamefont {V.~B.}\ \bibnamefont {Verma}}, \bibinfo {author}
  {\bibfnamefont {M.~J.}\ \bibnamefont {Stevens}}, \bibinfo {author}
  {\bibfnamefont {F.}~\bibnamefont {Marsili}}, \bibinfo {author} {\bibfnamefont
  {R.~P.}\ \bibnamefont {Mirin}}, \bibinfo {author} {\bibfnamefont {S.~W.}\
  \bibnamefont {Nam}}, \ and\ \bibinfo {author} {\bibfnamefont {A.~M.}\
  \bibnamefont {Wodtke}},\ }\bibfield  {title} {\enquote {\bibinfo {title}
  {Ultra-sensitive mid-infrared emission spectrometer with sub-ns temporal
  resolution},}\ }\href {\doibase 10.1364/OE.26.014859} {\bibfield  {journal}
  {\bibinfo  {journal} {Opt. Express}\ }\textbf {\bibinfo {volume} {26}},\
  \bibinfo {pages} {14859--14868} (\bibinfo {year} {2018})}\BibitemShut
  {NoStop}%
\bibitem [{\citenamefont {Taylor}\ \emph {et~al.}(2019)\citenamefont {Taylor},
  \citenamefont {Morozov}, \citenamefont {Gemmell}, \citenamefont
  {Erotokritou}, \citenamefont {Miki}, \citenamefont {Terai},\ and\
  \citenamefont {Hadfield}}]{Hadfield2p3umLIDAR}%
  \BibitemOpen
  \bibfield  {author} {\bibinfo {author} {\bibfnamefont {G.~G.}\ \bibnamefont
  {Taylor}}, \bibinfo {author} {\bibfnamefont {D.}~\bibnamefont {Morozov}},
  \bibinfo {author} {\bibfnamefont {N.~R.}\ \bibnamefont {Gemmell}}, \bibinfo
  {author} {\bibfnamefont {K.}~\bibnamefont {Erotokritou}}, \bibinfo {author}
  {\bibfnamefont {S.}~\bibnamefont {Miki}}, \bibinfo {author} {\bibfnamefont
  {H.}~\bibnamefont {Terai}}, \ and\ \bibinfo {author} {\bibfnamefont {R.~H.}\
  \bibnamefont {Hadfield}},\ }\bibfield  {title} {\enquote {\bibinfo {title}
  {Photon counting lidar at 2.3\&\#x00b5;m wavelength with superconducting
  nanowires},}\ }\href {\doibase 10.1364/OE.27.038147} {\bibfield  {journal}
  {\bibinfo  {journal} {Opt. Express}\ }\textbf {\bibinfo {volume} {27}},\
  \bibinfo {pages} {38147--38158} (\bibinfo {year} {2019})}\BibitemShut
  {NoStop}%
\bibitem [{\citenamefont {Korzh}\ \emph {et~al.}(2020)\citenamefont {Korzh},
  \citenamefont {Zhao}, \citenamefont {Allmaras}, \citenamefont {Autry},
  \citenamefont {Bersin}, \citenamefont {Beyer}, \citenamefont {Briggs},
  \citenamefont {Bumble}, \citenamefont {Colangelo}, \citenamefont {Crouch},
  \citenamefont {Dane}, \citenamefont {Gerrits}, \citenamefont {Lita},
  \citenamefont {Marsili}, \citenamefont {Moody}, \citenamefont {Pe{\~{n}}a},
  \citenamefont {Ramirez}, \citenamefont {Rezac}, \citenamefont {Sinclair},
  \citenamefont {Stevens}, \citenamefont {Velasco}, \citenamefont {Verma},
  \citenamefont {Wollman}, \citenamefont {Xie}, \citenamefont {Zhu},
  \citenamefont {Hale}, \citenamefont {Spiropulu}, \citenamefont {Silverman},
  \citenamefont {Mirin}, \citenamefont {Nam}, \citenamefont {Kozorezov},
  \citenamefont {Shaw},\ and\ \citenamefont {Berggren}}]{IntrinsicNbNJitter}%
  \BibitemOpen
  \bibfield  {author} {\bibinfo {author} {\bibfnamefont {B.}~\bibnamefont
  {Korzh}}, \bibinfo {author} {\bibfnamefont {Q.-Y.}\ \bibnamefont {Zhao}},
  \bibinfo {author} {\bibfnamefont {S.}~\bibnamefont {Allmaras}, \bibfnamefont
  {Jason P.and~Frasca}}, \bibinfo {author} {\bibfnamefont {T.~M.}\ \bibnamefont
  {Autry}}, \bibinfo {author} {\bibfnamefont {E.~A.}\ \bibnamefont {Bersin}},
  \bibinfo {author} {\bibfnamefont {A.~D.}\ \bibnamefont {Beyer}}, \bibinfo
  {author} {\bibfnamefont {R.~M.}\ \bibnamefont {Briggs}}, \bibinfo {author}
  {\bibfnamefont {B.}~\bibnamefont {Bumble}}, \bibinfo {author} {\bibfnamefont
  {M.}~\bibnamefont {Colangelo}}, \bibinfo {author} {\bibfnamefont {G.~M.}\
  \bibnamefont {Crouch}}, \bibinfo {author} {\bibfnamefont {A.~E.}\
  \bibnamefont {Dane}}, \bibinfo {author} {\bibfnamefont {T.}~\bibnamefont
  {Gerrits}}, \bibinfo {author} {\bibfnamefont {A.~E.}\ \bibnamefont {Lita}},
  \bibinfo {author} {\bibfnamefont {F.}~\bibnamefont {Marsili}}, \bibinfo
  {author} {\bibfnamefont {G.}~\bibnamefont {Moody}}, \bibinfo {author}
  {\bibfnamefont {C.}~\bibnamefont {Pe{\~{n}}a}}, \bibinfo {author}
  {\bibfnamefont {E.}~\bibnamefont {Ramirez}}, \bibinfo {author} {\bibfnamefont
  {J.~D.}\ \bibnamefont {Rezac}}, \bibinfo {author} {\bibfnamefont
  {N.}~\bibnamefont {Sinclair}}, \bibinfo {author} {\bibfnamefont {M.~J.}\
  \bibnamefont {Stevens}}, \bibinfo {author} {\bibfnamefont {A.~E.}\
  \bibnamefont {Velasco}}, \bibinfo {author} {\bibfnamefont {V.~B.}\
  \bibnamefont {Verma}}, \bibinfo {author} {\bibfnamefont {E.~E.}\ \bibnamefont
  {Wollman}}, \bibinfo {author} {\bibfnamefont {S.}~\bibnamefont {Xie}},
  \bibinfo {author} {\bibfnamefont {D.}~\bibnamefont {Zhu}}, \bibinfo {author}
  {\bibfnamefont {P.~D.}\ \bibnamefont {Hale}}, \bibinfo {author}
  {\bibfnamefont {M.}~\bibnamefont {Spiropulu}}, \bibinfo {author}
  {\bibfnamefont {K.~L.}\ \bibnamefont {Silverman}}, \bibinfo {author}
  {\bibfnamefont {R.~P.}\ \bibnamefont {Mirin}}, \bibinfo {author}
  {\bibfnamefont {S.~W.}\ \bibnamefont {Nam}}, \bibinfo {author} {\bibfnamefont
  {A.~G.}\ \bibnamefont {Kozorezov}}, \bibinfo {author} {\bibfnamefont {M.~D.}\
  \bibnamefont {Shaw}}, \ and\ \bibinfo {author} {\bibfnamefont {K.~K.}\
  \bibnamefont {Berggren}},\ }\bibfield  {title} {\enquote {\bibinfo {title}
  {Demonstration of sub-3 ps temporal resolution with a superconducting
  nanowire single-photon detector},}\ }\href {\doibase
  10.1038/s41566-020-0589-x} {\bibfield  {journal} {\bibinfo  {journal} {Nature
  Photonics}\ }\textbf {\bibinfo {volume} {14}},\ \bibinfo {pages} {250--255}
  (\bibinfo {year} {2020})}\BibitemShut {NoStop}%
\bibitem [{\citenamefont {Caloz}\ \emph {et~al.}(2019)\citenamefont {Caloz},
  \citenamefont {Korzh}, \citenamefont {Ramirez}, \citenamefont
  {Schönenberger}, \citenamefont {Warburton}, \citenamefont {Zbinden},
  \citenamefont {Shaw},\ and\ \citenamefont
  {Bussières}}]{IntrinsicMolyJitter}%
  \BibitemOpen
  \bibfield  {author} {\bibinfo {author} {\bibfnamefont {M.}~\bibnamefont
  {Caloz}}, \bibinfo {author} {\bibfnamefont {B.}~\bibnamefont {Korzh}},
  \bibinfo {author} {\bibfnamefont {E.}~\bibnamefont {Ramirez}}, \bibinfo
  {author} {\bibfnamefont {C.}~\bibnamefont {Schönenberger}}, \bibinfo
  {author} {\bibfnamefont {R.~J.}\ \bibnamefont {Warburton}}, \bibinfo {author}
  {\bibfnamefont {H.}~\bibnamefont {Zbinden}}, \bibinfo {author} {\bibfnamefont
  {M.~D.}\ \bibnamefont {Shaw}}, \ and\ \bibinfo {author} {\bibfnamefont
  {F.}~\bibnamefont {Bussières}},\ }\bibfield  {title} {\enquote {\bibinfo
  {title} {Intrinsically-limited timing jitter in molybdenum silicide
  superconducting nanowire single-photon detectors},}\ }\href {\doibase
  10.1063/1.5113748} {\bibfield  {journal} {\bibinfo  {journal} {Journal of
  Applied Physics}\ }\textbf {\bibinfo {volume} {126}},\ \bibinfo {pages}
  {164501} (\bibinfo {year} {2019})},\ \Eprint
  {http://arxiv.org/abs/https://doi.org/10.1063/1.5113748}
  {https://doi.org/10.1063/1.5113748} \BibitemShut {NoStop}%
\bibitem [{\citenamefont {Cahall}\ \emph {et~al.}(2017)\citenamefont {Cahall},
  \citenamefont {Nicolich}, \citenamefont {Islam}, \citenamefont {Lafyatis},
  \citenamefont {Miller}, \citenamefont {Gauthier},\ and\ \citenamefont
  {Kim}}]{OpticaPNR}%
  \BibitemOpen
  \bibfield  {author} {\bibinfo {author} {\bibfnamefont {C.}~\bibnamefont
  {Cahall}}, \bibinfo {author} {\bibfnamefont {K.~L.}\ \bibnamefont
  {Nicolich}}, \bibinfo {author} {\bibfnamefont {N.~T.}\ \bibnamefont {Islam}},
  \bibinfo {author} {\bibfnamefont {G.~P.}\ \bibnamefont {Lafyatis}}, \bibinfo
  {author} {\bibfnamefont {A.~J.}\ \bibnamefont {Miller}}, \bibinfo {author}
  {\bibfnamefont {D.~J.}\ \bibnamefont {Gauthier}}, \ and\ \bibinfo {author}
  {\bibfnamefont {J.}~\bibnamefont {Kim}},\ }\bibfield  {title} {\enquote
  {\bibinfo {title} {Multi-photon detection using a conventional
  superconducting nanowire single-photon detector},}\ }\href {\doibase
  10.1364/OPTICA.4.001534} {\bibfield  {journal} {\bibinfo  {journal} {Optica}\
  }\textbf {\bibinfo {volume} {4}},\ \bibinfo {pages} {1534--1535} (\bibinfo
  {year} {2017})}\BibitemShut {NoStop}%
\bibitem [{\citenamefont {Zhu}\ \emph {et~al.}(2020)\citenamefont {Zhu},
  \citenamefont {Colangelo}, \citenamefont {Chen}, \citenamefont {Korzh},
  \citenamefont {Wong}, \citenamefont {Shaw},\ and\ \citenamefont
  {Berggren}}]{STANDPNR}%
  \BibitemOpen
  \bibfield  {author} {\bibinfo {author} {\bibfnamefont {D.}~\bibnamefont
  {Zhu}}, \bibinfo {author} {\bibfnamefont {M.}~\bibnamefont {Colangelo}},
  \bibinfo {author} {\bibfnamefont {C.}~\bibnamefont {Chen}}, \bibinfo {author}
  {\bibfnamefont {B.~A.}\ \bibnamefont {Korzh}}, \bibinfo {author}
  {\bibfnamefont {F.~N.~C.}\ \bibnamefont {Wong}}, \bibinfo {author}
  {\bibfnamefont {M.~D.}\ \bibnamefont {Shaw}}, \ and\ \bibinfo {author}
  {\bibfnamefont {K.~K.}\ \bibnamefont {Berggren}},\ }\bibfield  {title}
  {\enquote {\bibinfo {title} {Resolving photon numbers using a superconducting
  nanowire with impedance-matching taper},}\ }\href {\doibase
  10.1021/acs.nanolett.0c00985} {\bibfield  {journal} {\bibinfo  {journal}
  {Nano Letters}\ }\textbf {\bibinfo {volume} {20}},\ \bibinfo {pages}
  {3858--3863} (\bibinfo {year} {2020})},\ \bibinfo {note} {pMID: 32271591},\
  \Eprint {http://arxiv.org/abs/https://doi.org/10.1021/acs.nanolett.0c00985}
  {https://doi.org/10.1021/acs.nanolett.0c00985} \BibitemShut {NoStop}%
\bibitem [{\citenamefont {Agne}\ \emph {et~al.}(2017)\citenamefont {Agne},
  \citenamefont {Kauten}, \citenamefont {Jin}, \citenamefont {Meyer-Scott},
  \citenamefont {Salvail}, \citenamefont {Hamel}, \citenamefont {Resch},
  \citenamefont {Weihs},\ and\ \citenamefont {Jennewein}}]{ThreePhoton}%
  \BibitemOpen
  \bibfield  {author} {\bibinfo {author} {\bibfnamefont {S.}~\bibnamefont
  {Agne}}, \bibinfo {author} {\bibfnamefont {T.}~\bibnamefont {Kauten}},
  \bibinfo {author} {\bibfnamefont {J.}~\bibnamefont {Jin}}, \bibinfo {author}
  {\bibfnamefont {E.}~\bibnamefont {Meyer-Scott}}, \bibinfo {author}
  {\bibfnamefont {J.~Z.}\ \bibnamefont {Salvail}}, \bibinfo {author}
  {\bibfnamefont {D.~R.}\ \bibnamefont {Hamel}}, \bibinfo {author}
  {\bibfnamefont {K.~J.}\ \bibnamefont {Resch}}, \bibinfo {author}
  {\bibfnamefont {G.}~\bibnamefont {Weihs}}, \ and\ \bibinfo {author}
  {\bibfnamefont {T.}~\bibnamefont {Jennewein}},\ }\bibfield  {title} {\enquote
  {\bibinfo {title} {Observation of genuine three-photon interference},}\
  }\href {\doibase 10.1103/PhysRevLett.118.153602} {\bibfield  {journal}
  {\bibinfo  {journal} {Phys. Rev. Lett.}\ }\textbf {\bibinfo {volume} {118}},\
  \bibinfo {pages} {153602} (\bibinfo {year} {2017})}\BibitemShut {NoStop}%
\bibitem [{\citenamefont {Wang}\ \emph {et~al.}(2019)\citenamefont {Wang},
  \citenamefont {Qin}, \citenamefont {Ding}, \citenamefont {Chen},
  \citenamefont {Chen}, \citenamefont {You}, \citenamefont {He}, \citenamefont
  {Jiang}, \citenamefont {You}, \citenamefont {Wang}, \citenamefont
  {Schneider}, \citenamefont {Renema}, \citenamefont {H\"ofling}, \citenamefont
  {Lu},\ and\ \citenamefont {Pan}}]{20ModeBosonSampling}%
  \BibitemOpen
  \bibfield  {author} {\bibinfo {author} {\bibfnamefont {H.}~\bibnamefont
  {Wang}}, \bibinfo {author} {\bibfnamefont {J.}~\bibnamefont {Qin}}, \bibinfo
  {author} {\bibfnamefont {X.}~\bibnamefont {Ding}}, \bibinfo {author}
  {\bibfnamefont {M.-C.}\ \bibnamefont {Chen}}, \bibinfo {author}
  {\bibfnamefont {S.}~\bibnamefont {Chen}}, \bibinfo {author} {\bibfnamefont
  {X.}~\bibnamefont {You}}, \bibinfo {author} {\bibfnamefont {Y.-M.}\
  \bibnamefont {He}}, \bibinfo {author} {\bibfnamefont {X.}~\bibnamefont
  {Jiang}}, \bibinfo {author} {\bibfnamefont {L.}~\bibnamefont {You}}, \bibinfo
  {author} {\bibfnamefont {Z.}~\bibnamefont {Wang}}, \bibinfo {author}
  {\bibfnamefont {C.}~\bibnamefont {Schneider}}, \bibinfo {author}
  {\bibfnamefont {J.~J.}\ \bibnamefont {Renema}}, \bibinfo {author}
  {\bibfnamefont {S.}~\bibnamefont {H\"ofling}}, \bibinfo {author}
  {\bibfnamefont {C.-Y.}\ \bibnamefont {Lu}}, \ and\ \bibinfo {author}
  {\bibfnamefont {J.-W.}\ \bibnamefont {Pan}},\ }\bibfield  {title} {\enquote
  {\bibinfo {title} {Boson sampling with 20 input photons and a 60-mode
  interferometer in a $1{0}^{14}$-dimensional hilbert space},}\ }\href
  {\doibase 10.1103/PhysRevLett.123.250503} {\bibfield  {journal} {\bibinfo
  {journal} {Phys. Rev. Lett.}\ }\textbf {\bibinfo {volume} {123}},\ \bibinfo
  {pages} {250503} (\bibinfo {year} {2019})}\BibitemShut {NoStop}%
\bibitem [{\citenamefont {Paudel}\ \emph {et~al.}(2019)\citenamefont {Paudel},
  \citenamefont {Wong}, \citenamefont {Goggin}, \citenamefont {Kwiat},
  \citenamefont {Bracker}, \citenamefont {Yakes}, \citenamefont {Gammon},\ and\
  \citenamefont {Steel}}]{KwiatQD}%
  \BibitemOpen
  \bibfield  {author} {\bibinfo {author} {\bibfnamefont {U.}~\bibnamefont
  {Paudel}}, \bibinfo {author} {\bibfnamefont {J.~J.}\ \bibnamefont {Wong}},
  \bibinfo {author} {\bibfnamefont {M.}~\bibnamefont {Goggin}}, \bibinfo
  {author} {\bibfnamefont {P.~G.}\ \bibnamefont {Kwiat}}, \bibinfo {author}
  {\bibfnamefont {A.~S.}\ \bibnamefont {Bracker}}, \bibinfo {author}
  {\bibfnamefont {M.}~\bibnamefont {Yakes}}, \bibinfo {author} {\bibfnamefont
  {D.}~\bibnamefont {Gammon}}, \ and\ \bibinfo {author} {\bibfnamefont {D.~G.}\
  \bibnamefont {Steel}},\ }\bibfield  {title} {\enquote {\bibinfo {title}
  {Direct excitation of a single quantum dot with cavity-spdc photons},}\
  }\href {\doibase 10.1364/OE.27.016308} {\bibfield  {journal} {\bibinfo
  {journal} {Opt. Express}\ }\textbf {\bibinfo {volume} {27}},\ \bibinfo
  {pages} {16308--16319} (\bibinfo {year} {2019})}\BibitemShut {NoStop}%
\bibitem [{\citenamefont {Islam}\ \emph {et~al.}(2017)\citenamefont {Islam},
  \citenamefont {Lim}, \citenamefont {Cahall}, \citenamefont {Kim},\ and\
  \citenamefont {Gauthier}}]{GauthierQKD}%
  \BibitemOpen
  \bibfield  {author} {\bibinfo {author} {\bibfnamefont {N.~T.}\ \bibnamefont
  {Islam}}, \bibinfo {author} {\bibfnamefont {C.~C.~W.}\ \bibnamefont {Lim}},
  \bibinfo {author} {\bibfnamefont {C.}~\bibnamefont {Cahall}}, \bibinfo
  {author} {\bibfnamefont {J.}~\bibnamefont {Kim}}, \ and\ \bibinfo {author}
  {\bibfnamefont {D.~J.}\ \bibnamefont {Gauthier}},\ }\bibfield  {title}
  {\enquote {\bibinfo {title} {Provably secure and high-rate quantum key
  distribution with time-bin qudits},}\ }\href {\doibase
  10.1126/sciadv.1701491} {\bibfield  {journal} {\bibinfo  {journal} {Science
  Advances}\ }\textbf {\bibinfo {volume} {3}} (\bibinfo {year} {2017}),\
  10.1126/sciadv.1701491},\ \Eprint
  {http://arxiv.org/abs/https://advances.sciencemag.org/content/3/11/e1701491.full.pdf}
  {https://advances.sciencemag.org/content/3/11/e1701491.full.pdf} \BibitemShut
  {NoStop}%
\bibitem [{\citenamefont {Atallah}\ \emph {et~al.}(2019)\citenamefont
  {Atallah}, \citenamefont {Sica}, \citenamefont {Shin}, \citenamefont
  {Friedman}, \citenamefont {Kahrobai},\ and\ \citenamefont
  {Caram}}]{AtallahSpectroscopy}%
  \BibitemOpen
  \bibfield  {author} {\bibinfo {author} {\bibfnamefont {T.~L.}\ \bibnamefont
  {Atallah}}, \bibinfo {author} {\bibfnamefont {A.~V.}\ \bibnamefont {Sica}},
  \bibinfo {author} {\bibfnamefont {A.~J.}\ \bibnamefont {Shin}}, \bibinfo
  {author} {\bibfnamefont {H.~C.}\ \bibnamefont {Friedman}}, \bibinfo {author}
  {\bibfnamefont {Y.~K.}\ \bibnamefont {Kahrobai}}, \ and\ \bibinfo {author}
  {\bibfnamefont {J.~R.}\ \bibnamefont {Caram}},\ }\bibfield  {title} {\enquote
  {\bibinfo {title} {Decay-associated fourier spectroscopy: Visible to
  shortwave infrared time-resolved photoluminescence spectra},}\ }\href
  {\doibase 10.1021/acs.jpca.9b04924} {\bibfield  {journal} {\bibinfo
  {journal} {The Journal of Physical Chemistry A}\ }\textbf {\bibinfo {volume}
  {123}},\ \bibinfo {pages} {6792--6798} (\bibinfo {year} {2019})}\BibitemShut
  {NoStop}%
\bibitem [{\citenamefont {Boroson}\ \emph {et~al.}(2014)\citenamefont
  {Boroson}, \citenamefont {Robinson}, \citenamefont {Murphy}, \citenamefont
  {Burianek}, \citenamefont {Khatri}, \citenamefont {Kovalik}, \citenamefont
  {Sodnik},\ and\ \citenamefont {Cornwell}}]{LLCDsummary}%
  \BibitemOpen
  \bibfield  {author} {\bibinfo {author} {\bibfnamefont {D.~M.}\ \bibnamefont
  {Boroson}}, \bibinfo {author} {\bibfnamefont {B.~S.}\ \bibnamefont
  {Robinson}}, \bibinfo {author} {\bibfnamefont {D.~V.}\ \bibnamefont
  {Murphy}}, \bibinfo {author} {\bibfnamefont {D.~A.}\ \bibnamefont
  {Burianek}}, \bibinfo {author} {\bibfnamefont {F.}~\bibnamefont {Khatri}},
  \bibinfo {author} {\bibfnamefont {J.~M.}\ \bibnamefont {Kovalik}}, \bibinfo
  {author} {\bibfnamefont {Z.}~\bibnamefont {Sodnik}}, \ and\ \bibinfo {author}
  {\bibfnamefont {D.~M.}\ \bibnamefont {Cornwell}},\ }\bibfield  {title}
  {\enquote {\bibinfo {title} {{Overview and results of the Lunar Laser
  Communication Demonstration}},}\ }in\ \href {\doibase 10.1117/12.2045508}
  {\emph {\bibinfo {booktitle} {Free-Space Laser Communication and Atmospheric
  Propagation XXVI}}},\ Vol.\ \bibinfo {volume} {8971},\ \bibinfo {editor}
  {edited by\ \bibinfo {editor} {\bibfnamefont {H.}~\bibnamefont {Hemmati}}\
  and\ \bibinfo {editor} {\bibfnamefont {D.~M.}\ \bibnamefont {Boroson}}},\
  \bibinfo {organization} {International Society for Optics and Photonics}\
  (\bibinfo  {publisher} {SPIE},\ \bibinfo {year} {2014})\ pp.\ \bibinfo
  {pages} {213 -- 223}\BibitemShut {NoStop}%
\bibitem [{\citenamefont {Vyhnalek}, \citenamefont {Tedder},\ and\
  \citenamefont {Nappier}(2018)}]{GRCsnspd}%
  \BibitemOpen
  \bibfield  {author} {\bibinfo {author} {\bibfnamefont {B.~E.}\ \bibnamefont
  {Vyhnalek}}, \bibinfo {author} {\bibfnamefont {S.~A.}\ \bibnamefont
  {Tedder}}, \ and\ \bibinfo {author} {\bibfnamefont {J.~M.}\ \bibnamefont
  {Nappier}},\ }\bibfield  {title} {\enquote {\bibinfo {title} {{Performance
  and characterization of a modular superconducting nanowire single photon
  detector system for space-to-Earth optical communications links}},}\ }in\
  \href {\doibase 10.1117/12.2290397} {\emph {\bibinfo {booktitle} {Free-Space
  Laser Communication and Atmospheric Propagation XXX}}},\ Vol.\ \bibinfo
  {volume} {10524},\ \bibinfo {editor} {edited by\ \bibinfo {editor}
  {\bibfnamefont {H.}~\bibnamefont {Hemmati}}\ and\ \bibinfo {editor}
  {\bibfnamefont {D.~M.}\ \bibnamefont {Boroson}}},\ \bibinfo {organization}
  {International Society for Optics and Photonics}\ (\bibinfo  {publisher}
  {SPIE},\ \bibinfo {year} {2018})\ pp.\ \bibinfo {pages} {369 --
  377}\BibitemShut {NoStop}%
\bibitem [{\citenamefont {{Yang}}\ \emph {et~al.}(2007)\citenamefont {{Yang}},
  \citenamefont {{Kerman}}, \citenamefont {{Dauler}}, \citenamefont {{Anant}},
  \citenamefont {{Rosfjord}},\ and\ \citenamefont {{Berggren}}}]{KermanET}%
  \BibitemOpen
  \bibfield  {author} {\bibinfo {author} {\bibfnamefont {J.~K.~W.}\
  \bibnamefont {{Yang}}}, \bibinfo {author} {\bibfnamefont {A.~J.}\
  \bibnamefont {{Kerman}}}, \bibinfo {author} {\bibfnamefont {E.~A.}\
  \bibnamefont {{Dauler}}}, \bibinfo {author} {\bibfnamefont {V.}~\bibnamefont
  {{Anant}}}, \bibinfo {author} {\bibfnamefont {K.~M.}\ \bibnamefont
  {{Rosfjord}}}, \ and\ \bibinfo {author} {\bibfnamefont {K.~K.}\ \bibnamefont
  {{Berggren}}},\ }\bibfield  {title} {\enquote {\bibinfo {title} {Modeling the
  electrical and thermal response of superconducting nanowire single-photon
  detectors},}\ }\href {\doibase 10.1109/TASC.2007.898660} {\bibfield
  {journal} {\bibinfo  {journal} {IEEE Transactions on Applied
  Superconductivity}\ }\textbf {\bibinfo {volume} {17}},\ \bibinfo {pages}
  {581--585} (\bibinfo {year} {2007})}\BibitemShut {NoStop}%
\bibitem [{\citenamefont {Rambo}, \citenamefont {Conover},\ and\ \citenamefont
  {Miller}(2019)}]{QOSPW}%
  \BibitemOpen
  \bibfield  {author} {\bibinfo {author} {\bibfnamefont {T.~M.}\ \bibnamefont
  {Rambo}}, \bibinfo {author} {\bibfnamefont {A.~R.}\ \bibnamefont {Conover}},
  \ and\ \bibinfo {author} {\bibfnamefont {A.~J.}\ \bibnamefont {Miller}},\
  }\bibfield  {title} {\enquote {\bibinfo {title} {{Two Billion Photon Per
  Second, One Photon at a Time}},}\ }in\ \href
  {https://spw2019.polimi.it/wp-content/uploads/2019/10/Book-of-abstract_on-line_v3.pdf}
  {\emph {\bibinfo {booktitle} {Single-Photon Workshop 2019 Book of
  Abstracts}}},\ \bibinfo {editor} {edited by\ \bibinfo {editor} {\bibfnamefont
  {H.}~\bibnamefont {Hemmati}}\ and\ \bibinfo {editor} {\bibfnamefont {D.~M.}\
  \bibnamefont {Boroson}}}\ (\bibinfo {year} {2019})\ p.\ \bibinfo {pages}
  {165}\BibitemShut {NoStop}%
\bibitem [{\citenamefont {Huang}\ \emph {et~al.}(2018)\citenamefont {Huang},
  \citenamefont {Zhang}, \citenamefont {You}, \citenamefont {Zhang},
  \citenamefont {Lv}, \citenamefont {Wang}, \citenamefont {Liu}, \citenamefont
  {Li},\ and\ \citenamefont {Wang}}]{Huang_2018}%
  \BibitemOpen
  \bibfield  {author} {\bibinfo {author} {\bibfnamefont {J.}~\bibnamefont
  {Huang}}, \bibinfo {author} {\bibfnamefont {W.}~\bibnamefont {Zhang}},
  \bibinfo {author} {\bibfnamefont {L.}~\bibnamefont {You}}, \bibinfo {author}
  {\bibfnamefont {C.}~\bibnamefont {Zhang}}, \bibinfo {author} {\bibfnamefont
  {C.}~\bibnamefont {Lv}}, \bibinfo {author} {\bibfnamefont {Y.}~\bibnamefont
  {Wang}}, \bibinfo {author} {\bibfnamefont {X.}~\bibnamefont {Liu}}, \bibinfo
  {author} {\bibfnamefont {H.}~\bibnamefont {Li}}, \ and\ \bibinfo {author}
  {\bibfnamefont {Z.}~\bibnamefont {Wang}},\ }\bibfield  {title} {\enquote
  {\bibinfo {title} {High speed superconducting nanowire single-photon detector
  with nine interleaved nanowires},}\ }\href {\doibase
  10.1088/1361-6668/aac180} {\bibfield  {journal} {\bibinfo  {journal}
  {Superconductor Science and Technology}\ }\textbf {\bibinfo {volume} {31}},\
  \bibinfo {pages} {074001} (\bibinfo {year} {2018})}\BibitemShut {NoStop}%
\bibitem [{\citenamefont {Wollman}\ \emph {et~al.}(2019)\citenamefont
  {Wollman}, \citenamefont {Verma}, \citenamefont {Lita}, \citenamefont {Farr},
  \citenamefont {Shaw}, \citenamefont {Mirin},\ and\ \citenamefont
  {Nam}}]{NISTkilopixel}%
  \BibitemOpen
  \bibfield  {author} {\bibinfo {author} {\bibfnamefont {E.~E.}\ \bibnamefont
  {Wollman}}, \bibinfo {author} {\bibfnamefont {V.~B.}\ \bibnamefont {Verma}},
  \bibinfo {author} {\bibfnamefont {A.~E.}\ \bibnamefont {Lita}}, \bibinfo
  {author} {\bibfnamefont {W.~H.}\ \bibnamefont {Farr}}, \bibinfo {author}
  {\bibfnamefont {M.~D.}\ \bibnamefont {Shaw}}, \bibinfo {author}
  {\bibfnamefont {R.~P.}\ \bibnamefont {Mirin}}, \ and\ \bibinfo {author}
  {\bibfnamefont {S.~W.}\ \bibnamefont {Nam}},\ }\bibfield  {title} {\enquote
  {\bibinfo {title} {Kilopixel array of superconducting nanowire single-photon
  detectors},}\ }\href {\doibase 10.1364/OE.27.035279} {\bibfield  {journal}
  {\bibinfo  {journal} {Opt. Express}\ }\textbf {\bibinfo {volume} {27}},\
  \bibinfo {pages} {35279--35289} (\bibinfo {year} {2019})}\BibitemShut
  {NoStop}%
\bibitem [{\citenamefont {Allmaras}\ \emph {et~al.}(2020)\citenamefont
  {Allmaras}, \citenamefont {Wollman}, \citenamefont {Beyer}, \citenamefont
  {Briggs}, \citenamefont {Korzh}, \citenamefont {Bumble},\ and\ \citenamefont
  {Shaw}}]{ThermalArray}%
  \BibitemOpen
  \bibfield  {author} {\bibinfo {author} {\bibfnamefont {J.~P.}\ \bibnamefont
  {Allmaras}}, \bibinfo {author} {\bibfnamefont {E.~E.}\ \bibnamefont
  {Wollman}}, \bibinfo {author} {\bibfnamefont {A.~D.}\ \bibnamefont {Beyer}},
  \bibinfo {author} {\bibfnamefont {R.~M.}\ \bibnamefont {Briggs}}, \bibinfo
  {author} {\bibfnamefont {B.~A.}\ \bibnamefont {Korzh}}, \bibinfo {author}
  {\bibfnamefont {B.}~\bibnamefont {Bumble}}, \ and\ \bibinfo {author}
  {\bibfnamefont {M.~D.}\ \bibnamefont {Shaw}},\ }\bibfield  {title} {\enquote
  {\bibinfo {title} {Demonstration of a thermally coupled row-column snspd
  imaging array},}\ }\href {\doibase 10.1021/acs.nanolett.0c00246} {\bibfield
  {journal} {\bibinfo  {journal} {Nano Letters}\ }\textbf {\bibinfo {volume}
  {20}},\ \bibinfo {pages} {2163--2168} (\bibinfo {year} {2020})},\ \bibinfo
  {note} {pMID: 32091221},\ \Eprint
  {http://arxiv.org/abs/https://doi.org/10.1021/acs.nanolett.0c00246}
  {https://doi.org/10.1021/acs.nanolett.0c00246} \BibitemShut {NoStop}%
\bibitem [{\citenamefont {Chiles}\ \emph {et~al.}(2020)\citenamefont {Chiles},
  \citenamefont {Buckley}, \citenamefont {Lita}, \citenamefont {Verma},
  \citenamefont {Allmaras}, \citenamefont {Korzh}, \citenamefont {Shaw},
  \citenamefont {Shainline}, \citenamefont {Mirin},\ and\ \citenamefont
  {Nam}}]{microwire}%
  \BibitemOpen
  \bibfield  {author} {\bibinfo {author} {\bibfnamefont {J.}~\bibnamefont
  {Chiles}}, \bibinfo {author} {\bibfnamefont {S.~M.}\ \bibnamefont {Buckley}},
  \bibinfo {author} {\bibfnamefont {A.}~\bibnamefont {Lita}}, \bibinfo {author}
  {\bibfnamefont {V.~B.}\ \bibnamefont {Verma}}, \bibinfo {author}
  {\bibfnamefont {J.}~\bibnamefont {Allmaras}}, \bibinfo {author}
  {\bibfnamefont {B.}~\bibnamefont {Korzh}}, \bibinfo {author} {\bibfnamefont
  {M.~D.}\ \bibnamefont {Shaw}}, \bibinfo {author} {\bibfnamefont {J.~M.}\
  \bibnamefont {Shainline}}, \bibinfo {author} {\bibfnamefont {R.~P.}\
  \bibnamefont {Mirin}}, \ and\ \bibinfo {author} {\bibfnamefont {S.~W.}\
  \bibnamefont {Nam}},\ }\bibfield  {title} {\enquote {\bibinfo {title}
  {Superconducting microwire detectors based on wsi with single-photon
  sensitivity in the near-infrared},}\ }\href {\doibase 10.1063/5.0006221}
  {\bibfield  {journal} {\bibinfo  {journal} {Applied Physics Letters}\
  }\textbf {\bibinfo {volume} {116}},\ \bibinfo {pages} {242602} (\bibinfo
  {year} {2020})},\ \Eprint
  {http://arxiv.org/abs/https://doi.org/10.1063/5.0006221}
  {https://doi.org/10.1063/5.0006221} \BibitemShut {NoStop}%
\bibitem [{\citenamefont {Miller}\ \emph {et~al.}(2011)\citenamefont {Miller},
  \citenamefont {Lita}, \citenamefont {Calkins}, \citenamefont {Vayshenker},
  \citenamefont {Gruber},\ and\ \citenamefont {Nam}}]{SelfAlignment}%
  \BibitemOpen
  \bibfield  {author} {\bibinfo {author} {\bibfnamefont {A.~J.}\ \bibnamefont
  {Miller}}, \bibinfo {author} {\bibfnamefont {A.~E.}\ \bibnamefont {Lita}},
  \bibinfo {author} {\bibfnamefont {B.}~\bibnamefont {Calkins}}, \bibinfo
  {author} {\bibfnamefont {I.}~\bibnamefont {Vayshenker}}, \bibinfo {author}
  {\bibfnamefont {S.~M.}\ \bibnamefont {Gruber}}, \ and\ \bibinfo {author}
  {\bibfnamefont {S.~W.}\ \bibnamefont {Nam}},\ }\bibfield  {title} {\enquote
  {\bibinfo {title} {Compact cryogenic self-aligning fiber-to-detector coupling
  with losses below one percent},}\ }\href {\doibase 10.1364/OE.19.009102}
  {\bibfield  {journal} {\bibinfo  {journal} {Opt. Express}\ }\textbf {\bibinfo
  {volume} {19}},\ \bibinfo {pages} {9102--9110} (\bibinfo {year}
  {2011})}\BibitemShut {NoStop}%
\bibitem [{\citenamefont {Rambo}\ and\ \citenamefont
  {Miller}(2019)}]{QOpatent}%
  \BibitemOpen
  \bibfield  {author} {\bibinfo {author} {\bibfnamefont {T.~M.}\ \bibnamefont
  {Rambo}}\ and\ \bibinfo {author} {\bibfnamefont {A.~J.}\ \bibnamefont
  {Miller}},\ }\href@noop {} {}\bibinfo {howpublished} {{U.S. Patent App. No.}
  16/399,207} (\bibinfo {year} {2019})\BibitemShut {NoStop}%
\bibitem [{\citenamefont {Vyhnalek}\ \emph {et~al.}(2019)\citenamefont
  {Vyhnalek}, \citenamefont {Tedder}, \citenamefont {Katz},\ and\ \citenamefont
  {Nappier}}]{GRCFMF}%
  \BibitemOpen
  \bibfield  {author} {\bibinfo {author} {\bibfnamefont {B.~E.}\ \bibnamefont
  {Vyhnalek}}, \bibinfo {author} {\bibfnamefont {S.~A.}\ \bibnamefont
  {Tedder}}, \bibinfo {author} {\bibfnamefont {E.~J.}\ \bibnamefont {Katz}}, \
  and\ \bibinfo {author} {\bibfnamefont {J.~M.}\ \bibnamefont {Nappier}},\
  }\bibfield  {title} {\enquote {\bibinfo {title} {{Few-mode fiber coupled
  superconducting nanowire single-photon detectors for photon efficient optical
  communications}},}\ }in\ \href {\doibase 10.1117/12.2510958} {\emph {\bibinfo
  {booktitle} {Free-Space Laser Communications XXXI}}},\ Vol.\ \bibinfo
  {volume} {10910},\ \bibinfo {editor} {edited by\ \bibinfo {editor}
  {\bibfnamefont {H.}~\bibnamefont {Hemmati}}\ and\ \bibinfo {editor}
  {\bibfnamefont {D.~M.}\ \bibnamefont {Boroson}}},\ \bibinfo {organization}
  {International Society for Optics and Photonics}\ (\bibinfo  {publisher}
  {SPIE},\ \bibinfo {year} {2019})\ pp.\ \bibinfo {pages} {62 --
  75}\BibitemShut {NoStop}%
\bibitem [{\citenamefont {{Zhang}}\ \emph {et~al.}(2019)\citenamefont
  {{Zhang}}, \citenamefont {{Zhang}}, \citenamefont {{You}}, \citenamefont
  {{Huang}}, \citenamefont {{Li}}, \citenamefont {{Sun}}, \citenamefont
  {{Wang}}, \citenamefont {{Lv}}, \citenamefont {{Zhou}}, \citenamefont
  {{Liu}}, \citenamefont {{Wang}},\ and\ \citenamefont
  {{Xie}}}]{MMsnspdFiltering}%
  \BibitemOpen
  \bibfield  {author} {\bibinfo {author} {\bibfnamefont {C.}~\bibnamefont
  {{Zhang}}}, \bibinfo {author} {\bibfnamefont {W.}~\bibnamefont {{Zhang}}},
  \bibinfo {author} {\bibfnamefont {L.}~\bibnamefont {{You}}}, \bibinfo
  {author} {\bibfnamefont {J.}~\bibnamefont {{Huang}}}, \bibinfo {author}
  {\bibfnamefont {H.}~\bibnamefont {{Li}}}, \bibinfo {author} {\bibfnamefont
  {X.}~\bibnamefont {{Sun}}}, \bibinfo {author} {\bibfnamefont
  {H.}~\bibnamefont {{Wang}}}, \bibinfo {author} {\bibfnamefont
  {C.}~\bibnamefont {{Lv}}}, \bibinfo {author} {\bibfnamefont {H.}~\bibnamefont
  {{Zhou}}}, \bibinfo {author} {\bibfnamefont {X.}~\bibnamefont {{Liu}}},
  \bibinfo {author} {\bibfnamefont {Z.}~\bibnamefont {{Wang}}}, \ and\ \bibinfo
  {author} {\bibfnamefont {X.}~\bibnamefont {{Xie}}},\ }\bibfield  {title}
  {\enquote {\bibinfo {title} {Suppressing dark counts of
  multimode-fiber-coupled superconducting nanowire single-photon detector},}\
  }\href {\doibase 10.1109/JPHOT.2019.2937537} {\bibfield  {journal} {\bibinfo
  {journal} {IEEE Photonics Journal}\ }\textbf {\bibinfo {volume} {11}},\
  \bibinfo {pages} {1--8} (\bibinfo {year} {2019})}\BibitemShut {NoStop}%
\bibitem [{\citenamefont {Reddy}\ \emph
  {et~al.}(2019{\natexlab{b}})\citenamefont {Reddy}, \citenamefont {Lita},
  \citenamefont {Nam}, \citenamefont {Mirin},\ and\ \citenamefont
  {Verma}}]{Reddy:19}%
  \BibitemOpen
  \bibfield  {author} {\bibinfo {author} {\bibfnamefont {D.~V.}\ \bibnamefont
  {Reddy}}, \bibinfo {author} {\bibfnamefont {A.~E.}\ \bibnamefont {Lita}},
  \bibinfo {author} {\bibfnamefont {S.~W.}\ \bibnamefont {Nam}}, \bibinfo
  {author} {\bibfnamefont {R.~P.}\ \bibnamefont {Mirin}}, \ and\ \bibinfo
  {author} {\bibfnamefont {V.~B.}\ \bibnamefont {Verma}},\ }\bibfield  {title}
  {\enquote {\bibinfo {title} {Achieving 98\% system efficiency at 1550 nm in
  superconducting nanowire single photon detectors},}\ }in\ \href {\doibase
  10.1364/CQO.2019.W2B.2} {\emph {\bibinfo {booktitle} {Rochester Conference on
  Coherence and Quantum Optics (CQO-11)}}}\ (\bibinfo  {publisher} {Optical
  Society of America},\ \bibinfo {year} {2019})\ p.\ \bibinfo {pages}
  {W2B.2}\BibitemShut {NoStop}%
\end{thebibliography}%

\end{document}